\documentclass[aps,prl,twocolumn,showpacs,amsmath,amssymb,superscriptaddress,floatfix]{revtex4}

\usepackage{bm} 
\usepackage{graphicx}
\usepackage{amssymb}
\usepackage{amsmath}
\usepackage{dcolumn} 
\usepackage[dvips]{color}
\usepackage{longtable}

\begin{document}

\title{
Quantum Hall Effect of Massless Dirac Fermions in a Vanishing Magnetic Field
      }

\date{\today}

\author{Kentaro Nomura}
\affiliation{
Department of Physics, Tohoku University, Sendai, 980-8578, Japan
            }

\author{Shinsei Ryu}
\affiliation{
Kavli Institute for Theoretical Physics
University of California, 
Santa Barbara, 
CA 93106, 
USA
            }

\author{Mikito Koshino}
\affiliation{
Department of Physics, Tokyo Institute of Technology, 
Meguro-ku, Tokyo 152-8551, Japan
            }
\affiliation{
Physics Department, Columbia University, New York, New York 10027, USA
            }

\author{Christopher Mudry}
\affiliation{
Condensed Matter Theory Group, Paul Scherrer Institute, CH-5232 Villigen PSI,
Switzerland 
            }
 	  
\author{Akira Furusaki}
\affiliation{
Condensed Matter Theory Laboratory, RIKEN, Wako, Saitama 351-0198, Japan
            }

\begin{abstract}
The effect of strong long-range
disorder on the quantization of the Hall conductivity
$\sigma_{xy}$
in graphene is studied numerically.
It is shown that increasing Landau-level mixing 
progressively destroys all
plateaus in $\sigma_{xy}$
except the plateaus at $\sigma_{xy}=\mp e^2/2h$
(per valley and per spin).
The critical state at the Dirac point 
is robust to strong disorder and belongs to
the universality class of the conventional plateau transitions
in the integer quantum Hall effect. 
We propose that the breaking of time-reversal symmetry by
ripples in graphene can realize this quantum critical point
in a vanishing magnetic field.
\end{abstract}


\maketitle

Graphene, an isolated layer of graphite \cite{Novoselov_2004},
displays a remarkable quantization of the Hall conductivity
$\sigma^{\ }_{xy}$
when subjected to a magnetic field.
As shown in Refs.~[\onlinecite{Novoselov_2005}]
and [\onlinecite{Zhang_2005}],
$\sigma^{\ }_{xy}$ 
is measured to be either a negative or a positive half-integer 
$n+1/2$ in units of $4e^{2}/h$.
This should be contrasted with the usual
integer quantum Hall effect (IQHE) for which
quantum Hall plateaus of the Hall conductance
in SiMOS-FET or in GaAs/AlGaAs heterojunctions 
occur at positive integer values $n$ in units
of $e^{2}/h$~\cite{QHE_review}.
In the vicinity of the two non-equivalent corners
$K$ and $K'$ of the first Brillouin zone of graphene, 
the non-interacting electronic dispersion is linear, 
i.e., at the so-called Dirac point
it realizes a noninteracting massless Dirac Hamiltonian.
As the Hall conductivity of a massive
two-component Dirac fermion in $(2+1)$-dimensional
space and time has long been known to be a half-integer
in units of $e^{2}/h$~\cite{Deser82,Ludwig_1994}, 
the observed quantization
of the Hall conductivity of graphene 
can be attributed to 4 independent flavors of 
two-component Dirac fermions
with the spin-degeneracy accounting for $2$ and
the valley-degeneracy accounting for another $2$ flavors.

This explanation fails to account for 
Anderson localization due to
disorder in graphene, as recently emphasized
in Ref.~\onlinecite{Mirlin07}.
Disorder can introduce both inter-valley
and intra-valley scattering between the Bloch states in
the valleys centered 
at $K$ and $K'$~\cite{Suzuura_2002,Sheng_2006,Koshino_2007}.
The physics of localization for the Bloch (Landau) states
of graphene is predicted to depend sensitively
on the nature of the disorder and, in particular,
on the range of the spatial correlations of the impurities. 
When disorder is short-ranged, inter-valley scattering has
large matrix elements.
In this case, all Bloch states are localized at zero
magnetic field~\cite{Suzuura_2002},
while the IQHE is wiped out by a sufficiently 
strong disorder~\cite{Sheng_2006}.
On the other hand, 
when disorder is sufficiently long-ranged, inter-valley
scattering is negligible, and
the metallic phase remains stable to disorder
if time-reversal symmetry is present%
~\cite{Ostrovsky_2007,Ryu_2007,Bardarson_2007,Nomura_2007}.

As for the conventional IQHE, 
the quantization of the Hall conductivity in graphene
requires the existence of at least one critical 
state in each impurity-broaden Landau level (LL) induced by
a magnetic field. 
The destruction of the quantum Hall plateaus
in the conventional IQHE with increasing disorder strength%
~\cite{Ando_1983,Sheng_1998} 
can be understood in terms of a critical state in a 
disorder-broaden LL
migrating (levitating) to higher filling fraction as soon as 
the disorder induces significant LL mixing%
~\cite{Laughlin_1984}.
The same phenomenon has been shown to be operative
for all impurity-broaden LLs in graphene
when the disorder is short-ranged~\cite{Sheng_2006}.

In this letter, we argue
on the basis of numerical calculation that,
when the disorder is sufficiently long-ranged,
all but one critical states in graphene undergo levitation
as the disorder strength increases. 
The exception is the state 
at the Dirac point that remains critical whatever 
the disorder strength is. Consequently, graphene with 
long-range disorder has the remarkable property 
that the quantum Hall plateaus
with $\sigma^{\ }_{xy}=\pm e^2/2h$
per spin and per valley
survive in the limit of strong disorder. In the language 
of the renormalization group (RG), the scaling flows of 
the longitudinal ($\sigma_{xx}$) and
transverse ($\sigma_{xy}$) conductivities
for graphene with strong infinite-range disorder
are determined by two attractive fixed points at 
$\sigma^{\ }_{xy}=\pm e^2/2h$ 
and one repulsive fixed point at $\sigma^{\ }_{xy}=0$.
For comparison, the corresponding phase
diagram for graphene with strong short-range disorder
is conventional in that it is characterized by a single
attractive fixed point, the insulating phase with
$\sigma^{\ }_{xx}=\sigma^{\ }_{xy}=0$ \cite{Sheng_2006,Koshino_2007}.
Of course, the range of the disorder is always finite
in a sample of graphene~\cite{Nomura_2006}. 
However, the characteristic scattering length induced
by short-range impurities could be longer
than the phase coherence length for some range of
temperature. 
Indeed it has been found, 
on a sample of graphene with a high-mobility 
($\mu\simeq 2\times 10^4\, \mathrm{cm^{2}/Vs}$), 
that the longitudinal conductivity at
the Dirac point
at zero magnetic field 
never falls with decreasing
temperature down to $10\,$mK~\cite{Tan_2007}; 
i.e., intervalley scattering is irrelevant 
in this range of temperature.

The honeycomb lattice of graphene has two atoms per unit cell on 
sites labeled $A$ and $B$. Linearization of the 
noninteracting lattice Hamiltonian in the clean limit,
ignoring the spin degrees of freedom, 
yields~\cite{Novoselov_2004}
\begin{equation}
\begin{split}
\mathcal{H}:=
\begin{pmatrix}
\mathcal{H}_{K}
&
0
\\
0
&
\mathcal{H}_{K'}
\end{pmatrix},
\quad
\mathcal{H}_{K}=
v_F^{} \bm{\sigma}\cdot
\left[
-
i\hbar\bm{\nabla} 
+
e\bm{A}
\right]
\end{split}
\label{dirac_hamiltonian1}
\end{equation}
where $\mathcal{H}_{K'}$ is the transpose of $\mathcal{H}_{K}$,
$v_F^{}$ is the Fermi velocity at the Dirac point,
and the Pauli matrices $\bm{\sigma}$ 
act on the sublattice degrees of freedom.   
In a finite magnetic field 
$\bm{B}=\bm{\nabla}\times\bm{A}=
(0,0,B)$
the spectrum of $\mathcal{H}_{K}$ consists of LLs 
with eigenvalues $E_n={\rm sgn}(n)\hbar\omega_0\sqrt{|n|}$ 
with $n\in\mathbb{Z}$~\cite{Dirac_qhe}.
The energy and length scales are set by $\omega_0=\sqrt{2}v_F/\ell_{B}$
and $\ell_{B}=\sqrt{\hbar/eB}$.
The filling fraction $\nu>0$ ($\nu<0$)
is the ratio of
the number of occupied (empty) states
above (below) the Dirac point to the LL degeneracy.

\begin{figure}[b]
\begin{center}
\includegraphics[width=0.4\textwidth]{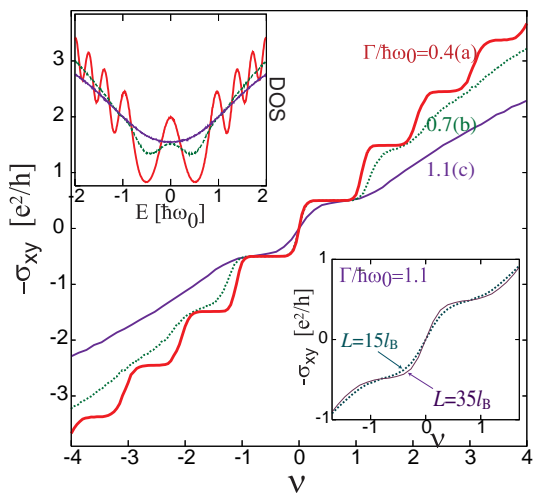}
\caption{(color online)
Disorder-average Hall conductivity $\sigma_{xy}$ as a function of the
filling fraction $\nu$ for the dimensionless disorder strength
(a) $\Gamma/\hbar\omega_0=0.4$, (b) 0.7, and (c) 1.1. 
Upper inset: Disorder-average density of states (DOS).
Lower inset: Size dependence of $\sigma_{xy}$ at $\Gamma/\hbar\omega_0=1.1$.
}
\label{Hall fct nu}
\end{center}
\end{figure}

In this work, as we neglect inter-valley and spin-dependent 
scattering channels, every LL is four-fold degenerate. 
The problem of Anderson localization that we shall consider is defined
by the random Dirac equation
$
\left[\mathcal{H}_{K}+V(\bm{r})\right]\psi=
\varepsilon\,\psi
$
with the static vector potential
in the Landau gauge $\bm{A}=(0,Bx,0)$
and $V$ a random static scalar potential.
We choose two-dimensional space to be a square with area $L^{2}$.
We introduce a cutoff $n_c=9$ for the LLs.
The disorder potential is
$V(\bm{r})=\sum_{j=1}^{N_{\mathrm{imp}}}
u_j 
\exp(-|\bm{r}-\bm{R}_j|^2/2d^2)/(2\pi d^2)
$
with $\bm{r}\in\mathbb{R}^{2}$,
$d$ a length scale, 
and $u/d^{2}=|u_j|/d^{2}$ an energy scale. 
The signs in
$u_j=\pm u$ 
and the scattering centers
$\bm{R}_j$
are chosen randomly and independently
at $N_{\mathrm{imp}}$ locations.
The disorder strength is estimated 
to be the broadening of the LLs
$
\Gamma^2/4=
2\pi u^2 N_{\mathrm{imp}}/[(\ell_{B}^2+2d^2)L^2]
$
within the self-consistent Born
approximation \cite{Ando_1983}.
In the following we set $d=0.7l_B$.

\begin{figure}[b]
\begin{center}
\includegraphics[width=0.4\textwidth]{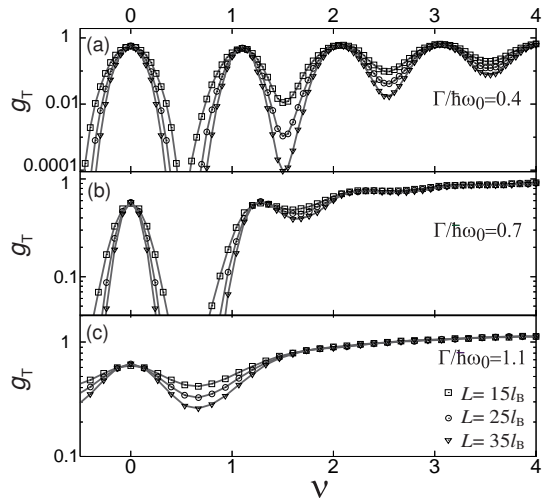}
\caption{
The Thouless conductance $g_T$ as a function
of the filling fraction $\nu$
when (a) $\Gamma/\hbar\omega_0=0.4$, (b) 0.7, and (c) 1.1,
for various system sizes,
$L/\ell_{B}=$ 15 ($\square$), 25 ($\circ$), and 35 ($\triangledown$).
        }
\label{Thouless fct nu}
\end{center}
\end{figure}

We evaluate the transverse conductivity with the Kubo formula
\begin{equation}
\sigma_{xy} =
-\frac{\hbar e^2}{L^2}\sum_{\varepsilon_m<\varepsilon_F<\varepsilon_n}
\overline{
\left(
\frac{{\rm Im}[\langle m|v_x|n\rangle\langle n|v_y|m\rangle]}
     {(\varepsilon_{m}-\varepsilon_{n})^2}
\right)
         },
\label{kubo}
\end{equation}
where $|n\rangle$ denotes an eigenstate with its
eigenvalue $\varepsilon_{n}$ of Hamiltonian $\mathcal{H}_K+V(\bm{r})$,
the corresponding  velocity operator is $\bm{v}:=v_F\bm{\sigma}$,
and the overline represents disorder averaging.
The dependence of the disorder-average Hall conductivity
$\sigma_{xy}$ on the filling
fraction is shown in Fig.~\ref{Hall fct nu}
for three values of the dimensionless disorder strength.
For small LL mixing induced by the disorder,
we observe that the curve (a) has
plateaus centered at 
half-integer values and with rounded corners
up to $\nu=\pm(n_c+1/2)$.
There are fewer well-defined Hall plateaus 
at moderately strong disorder strength
as shown in the curve (b).
Only the plateaus centered at $\nu=\pm1/2$ are seen
in the curve (c) when the disorder is strong
relative to $\hbar\omega_0$.
The size dependence of $\sigma_{xy}$ at strong disorder (c)
shown in the lower inset 
of Fig.~\ref{Hall fct nu}
is consistent with the expectation that 
$\sigma_{xy}$ has a step at $\nu=0$ 
in the $L\rightarrow\infty$ limit.
The quantization of $\sigma_{xy}/(e^2/h)$  
for all half-integers larger in magnitude than $1/2$
correlates with an oscillatory dependence on the energy of
the disorder-average density of states (DOS)
as is illustrated with the upper inset of
Fig.~\ref{Hall fct nu}.
This correlation is not present for the plateaus at $\nu=\pm1/2$.
LL mixing induced by the
disorder is thus responsible for the disappearance of
all Hall plateaus with $|\sigma_{xy}|>e^{2}/2h$
while the plateaus at
$\nu=\pm1/2$ are robust to strong disorder.

We now turn our attention to the dissipative 
component of electrical transport by 
studying the Thouless number
$
g_T\equiv \langle|\Delta\varepsilon|\rangle/\delta\varepsilon
$.
The Thouless energy 
$\Delta\varepsilon$
is the energy shift induced on energy eigenvalues
by using antiperiodic instead of periodic boundary
conditions \cite{Ando_1983}.
The mean level spacing is
$\delta\varepsilon=1/(L^2\overline{\rho(\varepsilon)})$ 
with $\rho(\varepsilon)$ the DOS.
The angular brackets denote the typical value
$\langle\Delta\varepsilon\rangle\equiv\exp(\overline{\ln\Delta\varepsilon})$.
The dependence of $g_T$ on $\nu$ at different values of
$L/\ell_{B}$
is shown in Fig.~\ref{Thouless fct nu}
for weak (a), moderate (b), and strong (c) 
disorder strength.
We observe a decrease of the Thouless number with the deviations
of the filling fraction from the integer values
$n$,
which is well correlated with the oscillatory dependence of the
DOS in Fig.~\ref{Hall fct nu}. 
This decrease gets more pronounced the closer $n$ is to $n=0$.
Local minima of the Thouless number are reached at half-integer
filling fractions. 
At fixed filling fraction,
$g_T$ generally decreases as the system size $L/\ell_{B}$
increases, while only at the center of an impurity-broaden LL it
is independent of $L/\ell_{B}$.
This is consistent with the existence of a
critical state close to integer filling fractions.
The decrease of $g_T$ gets more pronounced 
when approaching half-integer filling fractions.
Increasing disorder strength as shown in
Figs.~\ref{Thouless fct nu}(b) and~\ref{Thouless fct nu}(c) 
weakens the dependence of $g_T$ on both $\nu$ and
$L/\ell_{B}$
around all half-integer filling fractions larger than
$1/2$.
A decrease of $g_T$ remains clearly visible
at $\nu=1/2$ for strong disorder.

\begin{figure}
\begin{center}
\includegraphics[width=8.5cm,clip]{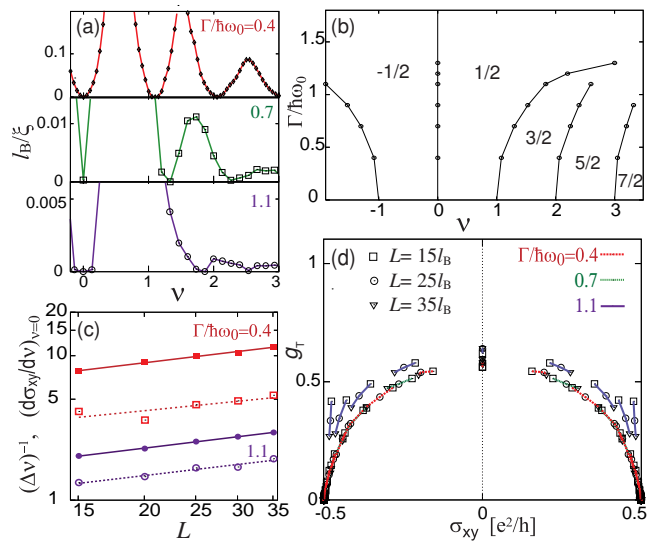} 
\caption{
(color online)
\label{fig: xi}
(a) Dependence on $\nu$ of the inverse localization length $1/\xi$.
(b) The phase diagram for the quantum Hall effect of massless Dirac fermions.
For each impurity broaden LL,
the phase boundaries are identified from the $L$-independent $g_T$ (circles).
Half-integers indicate 
$-\sigma_{xy}/(e^2/h)$.
(c) Widths of the half-maxima of the Thouless conductance (closed symbols) 
and the slope of the Hall conductivity (open symbols)
as functions of 
$L$.
(d) Scaling flow in $\sigma_{xy}-g_T$ plane:
$\sigma_{xy}$ and $g_T$
at $L/\ell_{B}=$ 15 ($\square$), 25 ($\circ$), and 35 ($\triangledown$) 
are plotted.
        }
\end{center}
\end{figure}

For any filling fraction,
a localization length $\xi$ is extracted from the fit 
$g_T(L)=g_0 \exp(-L/\xi)$ 
and $1/\xi$ is plotted in Fig.~\ref{fig: xi}(a)
for weak, moderate, and strong disorder.
Figure~\ref{fig: xi}(a) is consistent with
an oscillatory dependence of $\xi$ on $\nu$ that correlates with
the DOS in Fig.~\ref{Hall fct nu} 
and also shows that $\xi$ increases for $\nu>1$ with disorder strength
$\Gamma/(\hbar\omega_0)$.
The latter feature is familar in the conventional IQHE,
where the LL mixing makes the localization weaker
in the high-field regime \cite{Ando_1983}.
In each impurity-broadened LL
of Fig.~\ref{fig: xi}(a), there exists a $\nu$ for
which $\xi$
takes a maximum and much larger value than $L$.
This $\nu$ corresponds to a local maximum of $g_T$ 
and to a critical single-particle state.
In the limit of no LL mixing, 
critical states are found at integer values of $\nu$. 
Figure~\ref{fig: xi}(b) shows that all but one
critical states are increasing functions of $|\nu|$
as the amount of LL mixing induced by the disorder increases.
The exception is the $\nu=0$ critical state that remains 
pinned to the Dirac point for all disorder strength in  
Fig.~\ref{fig: xi}(b).
Consequently in the weak magnetic field limit 
$\hbar\omega_0/\Gamma\to0$
all single-particle states in the hole (electron) regime are localized:
$\sigma_{xy}/(e^2/h)=-1/2\,(+1/2)$
with the step discontinuity in $\sigma_{xy}$
occurring at $\nu=0$.
This behavior is different from the one for the conventional 
IQHE systems where $\sigma_{xy}\rightarrow 0$ 
(Hall insulator) 
in the weak magnetic field limit~\cite{Ando_1983,Sheng_1998}.
It is also different from the one observed in
numerical studies of the effects
of short-range correlated disorder on the IQHE in graphene%
~\cite{Sheng_2006},
which reported strong localization behavior in the vicinity of
$\nu=0$
over a wide range of magnetic field. This difference can be used
to distinguish whether the dominant source of disorder in graphene is 
short- or long-range. For $\nu$ close to $\nu=0$ and 
from weak to strong disorder,
Fig.~\ref{fig: xi}(c) 
is consistent with the scaling Ansatz
$\sigma_{xy}=f_{xy}(L^{1/\alpha_{\nu}}\nu)$
and
$g_{T}=f_{xx}(L^{1/\alpha_{\nu}}\nu)$
with $\alpha_{\nu}\approx2.3$.
Correspondingly,
$1/\xi\sim|\nu|^{\alpha_{\nu}}$,
$1/(\Delta_{T}\nu)\sim L^{\alpha_{\sigma}}$
where $\Delta_{T}\nu$ is the half-width of the peak
of $g_{T}$ at $\nu=0$,
and
$d\sigma_{xy}/d\nu\sim L^{\alpha_{\sigma}}$
with $\alpha_{\sigma}=1/\alpha_{\nu}$.
Moreover, 
$g_{T}=f_{xx}\biglb(f^{-1}_{xy}(\sigma_{xy})\bigrb)$
must hold close to $\nu=0$,
a prediction confirmed by 
the scaling flow in Fig.~\ref{fig: xi}(d),
where $(\sigma_{xy},g_T)$ computed at $L=15$, 25, and 35
are seen to move towards the attractive fixed points
at $(\pm e^2/2h,0)$ with increasing $L$.

\begin{figure}[t]
\begin{center}
\includegraphics[width=0.4\textwidth]{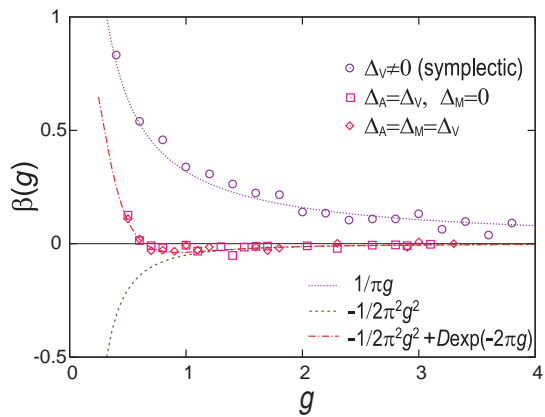}
\caption{(color online)
The $\beta$-functions for the random Dirac Hamiltonians%
~(\ref{dirac_hamiltonian2});
$\beta(g)$ is independent of the mean value of $V$.
        }
\label{beta fct}
\end{center}
\end{figure}

The phase boundary separating the quantum Hall phases
$\sigma_{xy}=-e^2/2h$ from $\sigma_{xy}=+e^2/2h$
in Fig.~\ref{fig: xi}(b) extends to the zero magnetic field limit 
$\hbar\omega_0/\Gamma\to0$.
This strongly suggests that the Dirac point is never localized
by long-range disorder, however strong the disorder is.
The Dirac point of graphene with long-range
disorder remains the critical point of the IQHE plateau
transition for any weak magnetic field.

We close this letter by studying the effects of 
effective time-reversal
symmetry breaking random perturbation on the zero-magnetic field
limit $\hbar\omega_0/\Gamma\to0$ 
of the Dirac Hamiltonian%
~(\ref{dirac_hamiltonian1}). The random single-particle Dirac Hamiltonian is
\cite{Ludwig_1994}
\begin{equation}
 \mathcal{H}:=
\bm{\sigma}\cdot[
-
i\hbar v_F\bm{\nabla}
+
\bm{a}(\bm{r})]
+
m(\bm{r})
\sigma_z+V(\bm{r}),
\label{dirac_hamiltonian2}
\end{equation}
where $\bm{a}(\bm{r})$,
$m(\bm{r})$, and 
$V(\bm{r})$ are Gaussian distributed with the covariances
$
\overline{
f_{\mu}(\bm{q})f_{\nu}(\bm{q}')
         }
=
\delta_{\mu\nu}\Delta_{\mu} \exp{(-\bm{q}^2d^2/2)}
\delta(\bm{q}+\bm{q}'),
$
if $(f_{\mu=0,1,2,3})=(V,a_1,a_2,m)$ whereby
$\Delta_{\mu}=\Delta_{V}$ for $\mu=0$,
$\Delta_{\mu}=\Delta_A$ for $\mu=1,2$,
and
$\Delta_{\mu}=\Delta_M$ for $\mu=3$.
The length scale $d$ is the range of the potential
that varies up to 1/30 of the minimal system size.
The ``random vector potential'' $\bm{a}(\bm{r})$
describes the effect of ripples in corrugated
graphene \cite{Morozov_2006,Morpurgo_2006}.
Ludwig \textit{et al.}\ in Ref.~\onlinecite{Ludwig_1994} 
conjectured that the random Dirac Hamiltonian 
(\ref{dirac_hamiltonian2})
flows into the critical point describing
the quantum Hall plateau transition.
Using the numerical method from Ref.~\cite{Nomura_2007}, 
we present in Fig.~\ref{beta fct}
the beta function $\beta(g)$ 
for the Kubo conductivity $g\equiv \sigma_{xx}/(e^2/h)$
which we fit against
$
 \beta(g)=-1/(2\pi^2g^2)+D\exp(-2\pi g)
$
\cite{QHE_review}
whenever time-reversal symmetry is broken
($-i\sigma_y\mathcal{H}^*i\sigma_y\ne\mathcal{H}$)
by either
$\Delta_A>0$ or $\Delta_M>0$.
The fitting parameter $D$ 
is optimized when $D=7$. Figure~\ref{beta fct}
is consistent with a fixed point at 
$g\simeq 0.6$ which is attractive in $g$.
The beta function in the limit $\Delta_{A,M}\to0$ with fixed
$\Delta_V>0$
is always positive 
(open circles in Fig.\ \ref{beta fct}),
as reported in Refs.~\cite{Bardarson_2007}
and~\cite{Nomura_2007}.
This limit is equivalent to the limit
$\hbar\omega_0/\Gamma\to0$ in Eq.~(\ref{dirac_hamiltonian1}).
Both limits induce a crossover from the unitary class with a topological
term to the symplectic class with a topological term%
~\cite{Ostrovsky_2007,Ryu_2007}.
This is different from the limit $\hbar\omega_0/\Gamma\to0$ 
in the conventional IQHE that realizes a crossover from the unitary class 
with a topological term to the orthogonal class.
The fact that the orthogonal class is always insulating
in two dimensions thus causes all critical states to levitate to large 
$\nu$ as $\hbar\omega_0/\Gamma\to0$
\cite{Ando_1983,Sheng_1998}.

We conclude by pointing out the interesting possibility
that ripples in graphene alone can realize scaling flows to
the IQHE plateau transition point
even in the absence of a magnetic field, i.e.,
graphene with long-range
disorder is very close to 
the quantum critical point of the IQHE.
This scenario is consistent
with recent low-temperature transport measurements
of high-mobility graphene samples
that imply a temperature independent minimal conductivity
\cite{Novoselov_2005,Tan_2007}.

This work was supported 
by the National Science Foundation under Grant No.\ PHY05-51164 and
by Grant-in-Aid for Scientific Research from MEXT (Grant No.\ 16GS0219).


\begin{thebibliography}{25}

\bibitem{Novoselov_2004} 
A. K. Geim and K. S. Novoselov,
Nature Mater. {\bf 6}, 183 (2007).

\bibitem{Novoselov_2005} 
K. S. Novoselov {\em et al.}, 
Nature (London) {\bf 438}, 197 (2005).

\bibitem{Zhang_2005} 
Y. Zhang {\em et al.}, 
Nature (London) {\bf 438}, 201 (2005). 

\bibitem{QHE_review}
{\it The Quantum Hall Effect}, 
edited by R. E. Prange and S. M. Girvin (Springer, New York, 1987).

\bibitem{Deser82}
S. Deser,
R. Jackiw, and S. Templeton,
Ann.\ Phys.\ (N.Y.) \textbf{140}, 372 (1982).

\bibitem{Ludwig_1994}
A. W. W. Ludwig \textit{et al.},
Phys.\ Rev.\ B \textbf{50}, 7526 (1994).

\bibitem{Mirlin07}
P. M. Ostrovsky,
I. V. Gornyi, and A. D. Mirlin,
arXiv:0712.0597.

\bibitem{Suzuura_2002} 
H. Suzuura and T. Ando, 
Phys.\ Rev.\ Lett.\ {\bf 89}, 266603 (2002);  
E. McCann \textit{et al}., \textit{ibid}.\ {\bf 97}, 146805 (2006); 
I. L. Aleiner and K. B. Efetov, \textit{ibid}.\ {\bf 97}, 236801 (2006);
A. Altland, \textit{ibid}.\ \textbf{97}, 236802 (2006).

\bibitem{Sheng_2006} 
D. N. Sheng, L. Sheng, and Z. Y. Weng, 
Phys.\ Rev.\ B {\bf 73}, 233406 (2006).

\bibitem{Koshino_2007} 
M. Koshino and T. Ando, 
Phys.\ Rev.\ B {\bf 75}, 033412 (2007).

\bibitem{Ostrovsky_2007}
P.\ M.\ Ostrovsky, I.\ V.\ Gornyi, and A.\ D.\ Mirlin,
Phys.\ Rev.\ Lett.\ \textbf{98}, 256801 (2007).

\bibitem{Ryu_2007}
S.\ Ryu, C.\ Mudry, H.\ Obuse, and A.\ Furusaki,
Phys.\ Rev.\ Lett.\ \textbf{99}, 116601 (2007).

\bibitem{Bardarson_2007} 
J. H. Bardarson, J. Tworzydlo, P. W. Brouwer, and C. W. J. Beenakker, 
Phys.\ Rev.\ Lett.\ {\bf 99}, 106801 (2007).

\bibitem{Nomura_2007} 
K. Nomura, M. Koshino, and S. Ryu, 
Phys.\ Rev.\ Lett.\ {\bf 99}, 146806 (2007).

\bibitem{Ando_1983} 
T. Ando, J.\ Phys.\ Soc.\ Jpn.\ {\bf 52}, 1740 (1983);
{\bf 53}, 3101 (1984); {\bf 53}, 3126 (1984); {\bf 55}, 3199 (1986).

\bibitem{Sheng_1998}
D. N. Sheng and Z. Y. Weng,
Phys.\ Rev.\ Lett.\ {\bf 80}, 580 (1998);
D. N. Sheng, Z. Y. Weng, and X. G. Wen,  
Phys.\ Rev.\ B {\bf 64}, 165317 (2001).

\bibitem{Laughlin_1984} 
D. E. Khmelnitskii, Phys.\ Lett.\ {\bf 106A}, 182 (1984); 
R. B. Laughlin, Phys.\ Rev.\ Lett.\ {\bf 52}, 2304 (1984).

\bibitem{Nomura_2006} 
K. Nomura and A. H. MacDonald, 
Phys.\ Rev.\ Lett.\ {\bf 96}, 256602 (2006); 
\textbf{98}, 076602 (2007);
T. Ando, J.\ Phys.\ Soc.\ Jpn.\ {\bf 75}, 074716 (2006).

\bibitem{Tan_2007}
Y. W. Tan \textit{et al.},
Eur.\ Phys.\ J.\ Special Topics \textbf{148}, 15 (2007).

\bibitem{Dirac_qhe}
J. W. McClure, Phys.\ Rev.\ \textbf{104}, 666 (1956).

\bibitem{Morozov_2006} 
S. V. Morozov \textit{et al}., 
Phys.\ Rev.\ Lett.\ {\bf 97}, 016801 (2006).

\bibitem{Morpurgo_2006}
A. F. Morpurgo and F. Guinea, 
Phys.\ Rev.\ Lett.\ \textbf{97}, 196804 (2006).

\end{thebibliography}
\end{document}